\documentclass[12pt]{iopart}

\usepackage{iopams}

\usepackage{epsfig}
\newcommand{\ga}{\alpha}
\newcommand{\gb}{\beta}
\newcommand{\gc}{\gamma}
\newcommand{\gC}{\Gamma}
\newcommand{\gd}{\delta}

\newcommand{\gve}{\varepsilon}
\newcommand{\gl}{\lambda}

\newcommand{\gL}{\Lambda}

\newcommand{\be}{\begin{equation}}
\newcommand{\ee}{\end{equation}}
\newcommand{\ba}{\begin{eqnarray}}
\newcommand{\ea}{\end{eqnarray}}
\newcommand{\as}[1]{\langle#1\rangle}
\newcommand{\ket}[1]{\vert #1 \rangle}
\newcommand{\hh}{\textstyle\frac{1}{2}}
\newcommand{\vk}[1]{\overline{#1}}

\begin{document}

\title[Light-front NJL model at finite temperature and density]{Light-front
Nambu--Jona-Lasinio model at finite temperature and density}

\author{S.~Strau\ss$^1$, S.~Mattiello$^2$, and M.~Beyer$^1$}
\address{$^1$ Institute of Physics, University of Rostock, D-18051 Rostock, Germany}
\address{$^2$ Institut f\"ur Theoretische Physik, Universit\"at Gie{\ss}en, D-35392 Giessen, Germany}
\ead{\mailto{stefan.strauss@uni-rostock.de},\mailto{stefano.mattiello@theo.physik.uni-giessen.de},
 \mailto{michael.beyer@uni-rostock.de}}

\begin{abstract}
In recent years light-front quantisation has been extended to allow
for a consistent treatment of systems at finite temperature and
density. This is in particular interesting for an investigation of the
processes in nuclear matter under extreme condition as occurring,
e.g., during a heavy ion collision. Utilising a Dyson expansion to the
$N$-point Green functions at finite temperature and density
we focus on the occurrence of pionic and scalar diquark
dynamics in quark matter and compute the masses and the Mott
dissociation using a separable $t$-matrix approach. For the scalar
quark-quark correlation we determine the critical temperature of colour
superconductivity using the Thouless criterion. On the same footing the
properties of the nucleon in a medium of quark matter are computed
within a Faddeev approach. Critical lines for nucleon breakup are
given. Presently, we use a light-front Nambu--Jona-Lasinio model
that allows us to compare these results of this novel approach to
the more traditional instant form approach, where
applicable.
\end{abstract}

%light cone quantisation \sep finite-temperature field theory \sep
%     quark-gluon plasma \sep  phase transitions

 %Quark deconfinement,
              %quark-gluon plasma production, and phase transitions
\vspace{2pc}
\submitto{\JPG}
\maketitle

\small
\normalsize

\section{Introduction}
Light-front quantisation recognized by Dirac~\cite{Dirac:cp} provides
a framework to describe the perturbative and the nonperturbative
regime of quantum chromodynamics (QCD), for an overview see
e.g.~\cite{Brodsky:1997de}. The importance of light-front quantisation
as a complement to Monte Carlo simulations has also recently been
emphasized by Ken Wilson~\cite{Wilson:2004de}.

A description involving both regimes is necessary, if one is
interested in the dynamics close to the chiral and the
confinement-deconfinement transition that is suggested by lattice
QCD~\cite{Karsch:2001vs,Karsch:2001cy,Allton:2003vx,Allton:2005gk,Cheng:2007jq},
model analyses~\cite{Alford:1997zt,Alford:1998mk,Rapp:1997zu,Rajagopal:2000wf},
and through recent interpretations of certain RHIC results
~\cite{Gyulassy:2004zy,Jacobs:2004qv,Stoecker:2004qu,Adcox:2004mh,Shuryak:2004cy,Heinz:2004pj,Song:2007fn}.
It seems that close to phase transition the quark gluon plasma does
not appear as a weakly interacting gas of quarks and gluons but rather
as a strongly correlated system.  In the framework of many-body Green
functions~\cite{kad62,fet71}, the inclusion of correlations can be
achieved by the Dyson equation approach developed in the context of
nonrelativistic nuclear physics, reviewed in~\cite{duk98}. This
approach has been generalised to light-front quantisation to comply
with the requirements of special relativity in~\cite{Beyer:2001bc},
where also the necessary ingredients such as distribution functions,
Green functions and Matsubara frequencies have been given. An early
exploration of finite temperature QED(1+1) in discrete light cone
quantisation has been done in Ref.~\cite{Elser:1996tq} and the
same theory has been reexamined in ~\cite{Strauss:2008zx}. Further
developments have been carried out and more examples given in Refs.
~\cite{Brodsky:2001ww,Alves:2002tx,Beyer:2003qb,Dalley:2004ca,Weldon:2003uz,Weldon:2003vh,Kvinikhidze:2003wc,Das:2003mf,Raufeisen:2004dg,Das:2005gm}.

A formal framework of covariant calculations at finite temperatures in
instant form has been given in
Refs.~\cite{Israel:1976tn,isr81,Weldon:aq} in a different
context. Combination of the covariant statistical physics and
light-front quantisation leads to the notion of general light-front
coordinates first introduced in~\cite{Weldon:2003uz}. These
coordinates allow the formulation of thermal field theory within the
light-front frame and help to overcome the feature that the
heat bath is moving with time-like velocity, while the time direction
vector is light-like. Choosing a certain set of parameters the oblique
frame in \cite{Weldon:2003uz} reduces to the general light-front frame
suggested in \cite{Beyer:2001bc,Alves:2002tx} which correspond to
a heat bath at rest in the instant form. The transformation is
repeated here for completeness~\cite{Weldon:2003uz}
\begin{eqnarray}
\label{eqn:generalLCframe}
x^{\bar 0}&=\:\;x^+, \qquad \qquad x_{\bar 0}&=\:\;x_0,\qquad \qquad
x_\perp=(x_1,x_2),\\ \nonumber
x^{\bar 3}&=\:\;x^3, \qquad \qquad x_{\bar 3}&=\:\;-2x_-, \qquad \quad
x^\perp=(x^1,x^2).
\end{eqnarray}
One recognizes that in (\ref{eqn:generalLCframe}) light cone and instant
form coordinates get mixed and the covariant time component is light cone time
while the contravariant component is ordinary time. The metric can be
derived by requiring invariance of the Minkowski line element and the
important condition $g_{\bar{0}\bar{0}}>0$ holds. Therefore the medium
velocity can be expressed as
$u^\mu=(1/\sqrt{g_{\bar{0}\bar{0}}},0,0,0)$ and fulfills $u^2=1$. Only
few applications of thermal field theory in the general light cone frame are
found in the literature.

Here we explore finite
temperature and density properties of few-body, namely one-, two- and
three-body systems in the Nambu--Jona-Lasinio (NJL) model that
reproduces some low energy phenomenology of QCD, in particular its
chiral properties, in a rather transparent way, and yet includes
spontaneous symmetry breaking as a nontrivial feature to be handled by
light-front quantisation. The investigation of the one- and two-body
sector confirms known results, like the phase boundary of chiral
restoration, the pion Mott transition, the onset of two flavour colour
superconductivity, from earlier calculations~\cite{Klevansky:qe,Buballa:2003qv} and
exemplifies the equivalence of thermal field theory in the front and
instant form. The three-body part consists of new results within this
approach.

This paper is organized as follows. In the second Section we derive
the explicit form of the meanfield thermal fermion propagator using
general light cone coordinates and thereby showing that one recovers
the propagator expected in the ordinary light-front frame. In Section
3 the propagator is embedded into the general framework of the Dyson
expansion of many-body Green functions. A systematic treatment for
two- and three-body correlations in the quark matter medium is
developed.  The last two Sections are concerned with applications
of the framework to correlations of two and three light quarks in
meanfield approximation. The Mott dissociation lines of the mesons and nucleons
are computed and plotted in the phase diagram of quark matter.

\section{Light-front equilibrium Green functions}
Let us summarise aspects of thermal field theory in the
light-front form and derive the free, in-medium Green function. In the
generalised coordinates (\ref{eqn:generalLCframe}) the product $u\cdot
P$ is given by $P_{\bar 0}/\sqrt{g_{\bar{0}\bar{0}}}$ and the grand
canonical partition function reads
\begin{equation}
Z_G =  \mathrm{Tr} \exp\left\{-(P_{\bar 0}/\sqrt{g_{\bar{0}\bar{0}}} -\mu N)/T\right\} =
\mathrm{Tr} \exp\left\{-(P_++P_- -\mu N)/T\right\},
\label{eqn:ZG}
\end{equation}
where $T$ is the scalar rest frame temperature and $\mu$ the chemical
potential belonging to the charge $N$. We introduce
$H_{\mathrm{eff}}\equiv P_++P_- -\mu N$ and the statistical operator
$\rho_Z=e^{-H_{\mathrm{eff}}/T}/Z_G$. The four-momentum operator and the charge
are
\begin{eqnarray}
P^\mu&=&\int d^2x^\perp dx^{\bar 3}\; T^{\bar{0}\mu}(x) = \int d^2x^\perp dx^3\; T^{+\mu}(x),
\label{eqn:P}\\
N&=&\int  d^2x^\perp dx^{\bar 3}\; J^{\bar 0}(x) = \int d^2x^\perp dx^3\; J^+(x),
\label{eqn:N}
\end{eqnarray}
where $T^{\mu\nu}(x)$ denotes the energy momentum tensor and
$J^\mu(x)$ the conserved current. Note that in
(\ref{eqn:P}) and (\ref{eqn:N}) the light cone densities are integrated over three
dimensional space and not the light-front plane.

The general light cone time-ordered (causal) Green function is
\begin{equation}
{\mathcal G}_{\ga\gb}(x-y)=
\theta(x^{\bar 0}-y^{\bar 0})\;{\mathcal G}^>_{\ga\gb}(x-y)
+ \theta(y^{\bar 0}-x^{\bar 0})\;{\mathcal G}^<_{\ga\gb}(x-y),
\label{eqn:defCrono}
\end{equation}
with the correlation functions
\begin{eqnarray}
{\mathcal G}^>_{\ga\gb}(x^{\bar 0}-y^{\bar 0},\bar{x}-\bar{y})&=&
-i\langle\Psi_\ga(x)\bar\Psi_\gb(y)\rangle,
\label{eqn:defGG}\\
{\mathcal G}^<_{\ga\gb}(x^{\bar 0}-y^{\bar 0},\bar{x}-\bar{y})&=&
\mp (-i)\langle\bar\Psi_\gb(y)\Psi_\ga(x)\rangle,
\label{eqn:defGL}
\end{eqnarray}
and the notation $\bar{x}=(x^{\bar 1},x^{\bar 2},x^{\bar 3})$.
Here and in the following the upper (lower) sign is for fermions (bosons).
The anti-causal Green function can be defined correspondingly.  In the
generalised Heisenberg picture the light cone time dependence of operators is
given by
\begin{equation}
{\mathcal O}(x^{\bar 0},\bar{x})=e^{iH_{\mathrm{eff}}x^{\bar 0}}{\mathcal O}(0,\bar{x})
e^{-iH_{\mathrm{eff}}x^{\bar 0}}.
\label{eqn:Heise}
\end{equation}
This differs from the regular Heisenberg picture of the vacuum by the
thermodynamic constraints.
In equilibrium the average is taken over the (equilibrium) grand
canonical statistical operator $\rho_G$, viz.
$\langle\cdots\rangle={\rm tr}\{\rho_G\dots\}$ and in addition
${\mathcal G}^<_{\alpha\beta}$ and ${\mathcal G}^>_{\alpha\beta}$ are
related by (anti) periodic boundary conditions,
\begin{equation}
{\mathcal G}_{\alpha\beta}^{<}(x^{\bar 0},\bar{x})=\pm {\mathcal G}_{\alpha\beta}^{>}(x^{\bar 0}-i\gb,\bar{x}).
\end{equation}
Therefore, in equilibrium, only one Green function needs to be considered,
and, alternatively to (\ref{eqn:defCrono}), we may introduce the thermodynamic
Green function~\cite{fet71}
\begin{equation}
{\mathcal G}_{\alpha\beta}^{\tau-\tau'}
=-\langle T_\tau \Psi_\ga(\tau)\bar\Psi_\gb(\tau')\rangle.
\label{eqn:thermo}
\end{equation}
Setting $x^+=-i\tau$ (``imaginary time'') in (\ref{eqn:Heise}) we achieve
the thermodynamic Heisenberg picture.  The Green functions
(\ref{eqn:defCrono}) and (\ref{eqn:thermo}) are related by their spectral
representation utilising analytic continuation of the respective spectral
functions just as in the instant form case~\cite{fet71}. To be more specific,
in momentum space we introduce the spectral function $A(k_{\bar 0},\underline{k})$
\begin{eqnarray}
{\mathcal G}^{<}(k_{\bar 0},\underline{k}) &=& f(k_{\bar 0},\underline{k}) A(k_{\bar 0},\underline{k})
\label{eqn:GL}\\
{\mathcal G}^{>}(k_{\bar 0},\underline{k}) &=& (1-f(k_{\bar 0},\underline{k})) A(k_{\bar 0},\underline{k})
\label{eqn:GG}
\end{eqnarray}
where $A(k_{\bar 0},\underline{k})=i({\mathcal G}^{>}(k_{\bar 0},\underline{k})-
{\mathcal G}^{<}(k_{\bar 0},\underline{k}))$
and $\underline{k}$ abbreviates $(k_{\bar 1},k_{\bar 2},
k_{\bar 3})$.
Using (\ref{eqn:defCrono}) along with (\ref{eqn:GL}) and (\ref{eqn:GG}) the
causal Green function can be represented as
\begin{equation}
{\mathcal  G}(k_{\bar 0},\underline{k})=\int\limits_{-\infty}^\infty
  \frac{d \tilde{k}_{\bar 0}}{2(2\pi)} \left[ \frac{f(\tilde{k}_{\bar 0},\underline{k})
A(\tilde{k}_{\bar 0},\underline{k})}{\hh k_{\bar 0} -\hh \tilde{k}_{\bar 0} - i\gve} +
    \frac{(1-f(\tilde{k}_{\bar 0},\underline{k})) A(\tilde{k}_{\bar 0},\underline{k})}{\hh k_{\bar 0}
-\hh \tilde{k}_{\bar 0} + i\gve}\right].
\label{eqn:Gk}
\end{equation}
A similar equation holds for (\ref{eqn:thermo}), where the $i\varepsilon$ term can be dropped
\begin{equation}
 {\mathcal G}(k^n_{\bar 0},\underline{k})=\int\limits_{-\infty}^\infty
  \frac{d \tilde{k}_{\bar 0}}{2(2\pi) }
  \frac{ A(\tilde{k}_{\bar 0},\underline{k})}{\hh k^n_{\bar 0} -\hh \tilde{k}_{\bar 0}}
\label{eqn:Gkthermo}
\end{equation}
and the Matsubara frequencies are given by
\begin{equation}
\hh k^n_{\bar 0} = \left\{\begin{array}{ll} i(2n+1)\pi T +\mu &{\rm fermions,}\\[1ex]
                            i2n\pi T +\mu &{\rm bosons.}\end{array}\right.
\end{equation}
One may introduce traditional light cone coordinates via (\ref{eqn:generalLCframe})
into the equations (\ref{eqn:GL}) to (\ref{eqn:Gkthermo}), since they only depend on the
difference $k_{\bar0}-\tilde{k}_{\bar 0}$.
Then the Matsubara frequencies in $k^-$ read
\begin{equation}
\hh k^n_- = \left\{\begin{array}{ll} i(2n+1)\pi T +\mu -\hh k^+&{\rm fermions,}\\[1ex]
                            i2n\pi T +\mu -\hh k^+&{\rm bosons,}\end{array}\right.
\end{equation}
and have been given before in Ref.~\cite{Beyer:2001bc} without the
introduction of the general light-front frame.
% In the imaginary time formalism the spectral function is
%\begin{equation}
%A(k)=
%2\pi i\lim_{\gve\rightarrow  0}\left(G()|_{k^-_n=k^-+i\gve}-G(k_n)|_{k^-_n=k^--i\gve}\right),
%\label{eqn:spectralfunction}
%\end{equation}
%Direct evaluation of (\ref{eqn:thermo}) with free fields leads
%to the Matsubara-Fourier representation of the free thermodynamic Green function
%\begin{equation}
%{\mathcal G}(k^n_{\bar 0},\underline{k})= \frac{\gamma k_{\mathrm{on}}+m}{k_n^2-m^2}.
%\label{eqn:Gim}
%\end{equation}
%Here $k_{\mathrm{on}}$ is the on-shell four-momentum.
For an ideal gas (and also in Hartree-Fock approximation utilised
later on) the spectral function is
\begin{eqnarray}
\nonumber
A(k_{\bar 0},\underline{k})&=& 2\pi\left(\gamma k +m\right)\epsilon(k_{\bar 0})\gd(k^2-m^2)\\
&=&2\pi\frac{\gamma k + m}{-2k_{\bar 3}}\;\epsilon(k_{\bar 3})\gd(k_{\bar 0} -  k_{\bar 0,\mathrm{on}}),
\label{eqn:spec}
\end{eqnarray}
where $\epsilon(x)$ denotes the sign function.
Inserting (\ref{eqn:spec}) into (\ref{eqn:Gk}) and transforming from general light cone 
frame back to the traditional one leads to
\begin{eqnarray}
{\mathcal G}(k)&=&
\frac{\gamma k_\mathrm{on}+m}{k^+}\epsilon(k^+)
\left(\frac{f(k^+,k_\perp)}
{k^-- k^-_\mathrm{on}-i\varepsilon}
+\frac{1-f(k^+,k_\perp)}{k^-- k^-_\mathrm{on}+i\varepsilon}\right),
\label{eqn:lfmed_short}
\end{eqnarray}
with $k^-_\mathrm{on}=(\vec k_\perp^2 +m^2)/k^+$. Separating positive
from negative $k^+$ components and introducing the grand canonical
Fermi distribution functions of particles $f^+\equiv f$ and antiparticles
$f^-(k^+)=1-f^+(-k^+)$
\begin{equation}
f^\pm(k^+,\vec k_\perp)=
\left[\exp\left\{\frac{1}{T}\left(\frac{1}{2}k^-_{\mathrm{on}}
+\frac{1}{2}k^+\mp\mu\right)\right\}+1\right]^{-1}
\label{eqn:fermipm}
\end{equation}
Eq.~ (\ref{eqn:lfmed_short}) changes to
\begin{eqnarray}
{\mathcal G}(k)&=&
\frac{\gamma k_\mathrm{on}+m}{k^-- k^-_\mathrm{on}+i\varepsilon}
 \frac{\theta (k^+)}{k^+}
(1-f^+)
+\frac{\gamma k_\mathrm{on}+m}{k^-- k^-_\mathrm{on}-i\varepsilon}
\frac{\theta(k^+)}{k^+}
f^+\label{eqn:lfmed_long}\\
&&
+\frac{\gamma k_\mathrm{on}+m}{k^-- k^-_\mathrm{on}+i\varepsilon}
\frac{\theta(-k^+)}{k^+}
f^-
+\frac{\gamma k_\mathrm{on}+m}{k^-- k^-_\mathrm{on}-i\varepsilon}
\frac{\theta(-k^+)}{k^+}
(1-f^-),
\nonumber
\end{eqnarray}
which has been given before in Ref.~\cite{Beyer:2005rd} following
naively a direct approach to light-front Green functions. However, the
results are as shown equal and the particle propagator, that is
setting $f^-=0$ in (\ref{eqn:lfmed_long}), can be found in
Ref.~\cite{Beyer:2001bc}. In comparison to (\ref{eqn:lfmed_long}) the
Green function given in Ref.~\cite{Alves:2002tx,Kvinikhidze:2003wc}
were derived using the closed time path formalism and have therefore
$2\times 2$ matrix structure. These Green function are suited to
non-equilibrium situations, while we here concentrate on the
equilibrium or close-to-equilibrium systems, which is why
(\ref{eqn:lfmed_long}) is sufficient.

%************************************************************************************************

\section[Dyson expansion]{Dyson expansion}\label{Abs:Dyson}
To investigate bound states in hot and dense quark matter we use
techniques of the many-body Green functions organizing the equations
into a Dyson expansion that leads to a hierarchy of linked cluster
equations.  This approach for the investigation of correlations in
many-body systems at finite temperature and density was derived by
P. Schuck and collaborators~\cite{duk98}.  Furthermore this formalism
was extended to the light-front in~\cite{Beyer:2001bc} to investigate
three-quark correlations in quark matter.  In the following we present
a covariant derivation of the Dyson approach on the light-front and
its application for mesonic and baryonic bound states.  The Dyson
approach to many-body Green functions in the light-front quantisation
allows to calculate systematically the properties of few-body
clusters, in particular of the two-quark bound states (viz. pion) and
of the three-quark clusters (viz. nucleon).  This sets up the
framework to investigate the change from nuclear to quark matter.  One
starts from the generalisation of the casual Green function given by
\ba
G^{x^{\bar 0}-x'^{\bar 0}}_{\alpha\beta} \!&=&\! -i\as{
    T_{x^{\bar 0}}A_{\alpha}(x^{\bar 0})A^{\dag}_{\beta}(x'^{\bar 0})} {}\\
&=&\!-i\theta(x^{\bar 0}-x'^{\bar 0})\as{A_{\alpha}(x^{\bar 0})A^{\dag}_{\beta}(x'^{\bar 0})}\pm
i\theta(x'^{\bar 0}-x^{\bar 0})\as{A^{\dag}_{\beta}(x'^{\bar 0})
A_{\alpha}(x^{\bar 0})}\nonumber .
\ea
The operators $A_\ga(x^{\bar 0})$ can be build out of any number of
field operators (fermions and/or bosons) and their light cone time
dependence in the Heisenberg picture follows Eq.~(\ref{eqn:Heise}).
For the free fermion field, i.e. $A(x^{\bar 0})=\Psi(x^{\bar 0})$, the
standard light-front propagator (\ref{eqn:defCrono}) is recovered.

>From the definition of the Green function at finite temperature we can derive the Dyson
equation
\be\label{eqn:BWG1}
i\partial_{\bar 0}G^{x^{\bar 0}-x'^{\bar 0}}_{\alpha\beta}=\delta(x^{\bar 0}-x'^{\bar 0})N^{x^{\bar 0}}_{\alpha\beta}+R^{x^{\bar 0}-x'^{\bar 0}}_{\alpha\beta},
\ee
where $\partial_{{\bar 0}}$ is the derivation with respect to $x^{\bar 0}$ and
\be
\label{def:N-R}
N^{x^{\bar 0}}_{\alpha\beta}=\as{\lbrack A_{\alpha},A^{\dag}_{\beta}\rbrack_{\pm}
(x^{\bar 0})} ,
\qquad R^{x^{\bar 0}-x'^{\bar 0}}_{\alpha\beta}=-i\as{T_{x^{\bar 0}} \lbrack
  A_{\alpha},H\rbrack(x^{\bar 0})A^{\dag}_{\beta}(x'^{\bar 0})} .
\ee
%To reduce the Dyson equation (\ref{eqn:BWG1}) in integral form we can
%introduce the inverse Green function \({\cal G}^{-1}\)
%\be\label{def:InvG}
%\sum_{\beta'}\int d\bar{x}^{\bar 0}\left({\cal
%    G}^{-1}\right)^{x_{1}^{\bar 0}-\bar{x}^{\bar 0}}_{\alpha'\beta'}{\cal
%  G}^{\bar{x}^{\bar 0}-x_{2}^{\bar 0}}_{\beta'\beta}=\delta(x_{1}^{\bar 0}-x_{2}^{\bar 0})
%\delta_{\alpha'\beta}.
%\ee
With the following definition of the mass matrix
\be\label{def:MassOp}
M^{x^{\bar 0}-\bar{x}^{\bar 0}}_{\alpha\beta'}=\sum_{\alpha'}\int dx^{\bar 0}R^{x^{\bar 0}-x_{1}^{\bar 0}}_{\alpha\alpha'}
\left(G^{-1}\right)^{x_{1}^{\bar 0}-\bar{x}^{\bar 0}}_{\alpha'\beta'}
\ee
the Dyson equation (\ref{eqn:BWG1}) can be written as
\be\label{eqn:Dyson}
i\partial_{\bar 0}G^{x^{\bar 0}-x'^{\bar 0}}_{\alpha\beta}=\delta(x^{\bar 0}-x'^{\bar 0})N^{x^{\bar 0}}_{\alpha\beta}
+\sum_{\beta'}\int d\bar{x}^{\bar 0} M^{x^{\bar 0}-\bar{x}^{\bar 0}}_{\alpha\beta'}G^{\bar{x}^{\bar 0}-x'^{\bar 0}}_{\beta'\beta} .
\ee
The mass matrix describes the modification of the particles energy and
of the interaction due to the medium.  The expression
(\ref{def:MassOp}) for the mass operator is not well suited for practical computations.
However, after some formal manipulations, the mass operator reads
\ba
M^{x^{\bar 0}-\bar{x}^{\bar 0}}_{\alpha\beta'}&=&\sum_{\alpha'}\lbrack
\delta(x^{\bar 0}-\bar{x}^{\bar 0})\as{\lbrack\lbrack A_{\alpha},H\rbrack,
A^{\dag}_{\alpha'}\rbrack_{\pm}\left(x^+\right)}{}\nonumber\\
& &{}-i\as{T_x^{\bar 0}\lbrack A_{\alpha},
H\rbrack(x^+)\lbrack H,A^{\dag}_{\alpha'}\rbrack
(x_1^{\bar 0})}_{\mathrm{irr.}}\rbrack\left(N^{-1}\right)^{x^{\bar 0}}_{\alpha'\beta'} .
\ea
This equation separates the mass matrix in a instantaneous term
$M_{0,\alpha\beta'}$ related to the meanfield approximation
and in a retardation (or memory) term $ M^{x^{\bar
0}-\bar{x}^{\bar 0}}_{\mathrm{r},\alpha\beta'}$, i.e.
\be
M^{x^{\bar 0}-\bar{x}^{\bar 0}}_{\alpha\beta'}=\delta(x^{\bar 0}-\bar{x}^{\bar 0})M_{0,\alpha\beta'}+
M^{x^{\bar 0}-\bar{x}^{\bar 0}}_{\mathrm{r},\alpha\beta'} .
\ee
Here, we neglect the retarded part of the mass operator which leads
to the following equation of motion
\be\label{eqn:Dyson-Matsubara}
\left(i\partial_{x^{\bar 0}}-M^{x^{\bar 0}}_0\right)G^{x^{\bar 0}-x'^{\bar 0}}=
\delta\left(x^{\bar 0}-x'^{\bar 0}\right)N^{x^{\bar 0}},
\ee
where the indices $\ga,\gb$ are suppressed for simplicity.

For an explicit calculation of the expression for $M_{0}$ and
the normalisation factor $N$ assumptions about the medium are
needed. Here, we assume homogeneous matter of non-interacting quarks
and antiquarks. In this quasi-particle approximation the calculation
of the mass operator and of the normalisation factor is performed in
momentum space by introducing creations operators for the particles
$b^\dag(\vk{k},s)$ and antiparticles $d^\dag(\vk{k},s)$.  The
assumptions imply directly the Fermi distribution functions
$f^\pm(\vk{k})$.  For the one-body case one obtains the in-medium propagator
(\ref{eqn:in-mediumgapeqn}) from the Dyson expansion. To investigate
mesonic bound states we have to calculate the
normalisation factor of the quark-antiquark system using~(\ref{def:N-R}). It reads
\be
\label{eqn:NormPion}
N_{2}(k_1,k_2)=%L_{\pi}(1,2)\left(
1-f^+_1(k_1)-f^-_2(k_2)
%\right)
\ee
and corresponds to the well-known Pauli blocking factor. Besides
mesons three-quark bound states like the nucleon are subject of this
paper, where we neglect the antiparticles degree of freedom. This is
justified in light-front dynamics because pair creation processes
are likely to be suppressed.  The normalisation factor of the
two-quark system is given by
\be\label{eqn:NormDiq}
N_{2}(k_1,k_2)=%L_{\pi}(1,2)\left(
1-f_{1}-f_{2}%\right)
,
\ee
where $f_{i}=f^+_{i}(k_i)$ for simplicity.
The corresponding factor of the three-quark system is
\be\label{eqn:Norm3B}
N_{3}(k_1,k_2,k_3)=%L_{\pi}(1,2)\left(
\bar{f}_{1}\bar{f}_{2}\bar{f}_{3}-f_{1}f_{2}f_{3}%\right)
,
\ee
where we used the notation $\bar{f}_{i}=1-f_{i}$. In general, the mass
operator \(M_{0}\) can be separated into two terms: the first term,
\(H_{0}\), contains the self energy corrections in the effective mass and
the second term describes the in-medium modified interaction.  In the
$t$-matrix equation we neglect the self energy term and
consider in the following the term \(H_{0}\) only.

The resolvent operator \(R^{(n)}_{0}\) can be defined by
\ba
%R^{(n)}\left(z\right) &=& \frac{1}{z-{\cal M}^{(0)}},\qquad
%z\notin\sigma\left({\cal M}^{(0)}\right)\\
R^{(n)}_{0}\left(z\right) &=& \frac{1}{z-H_{0}},\qquad
z\notin\sigma\left(H_{0}\right) ,
\ea
where $n=2,3$ indicates the two- and three-quark correlations respectively.
Hence the Dyson equation (\ref{eqn:Dyson-Matsubara}) yields
\ba\label{eqn:Gn}
G^{(n)}_{0}\left(z\right) &=& R^{(n)}_{0}\left(z\right)N_n,
\ea

In this way the dominant medium effects due to Pauli blocking and self
energy corrections are systematically included in the relativistic
equations for the bound states, because the $t$-matrix equations
contain $G^{(n)}_{0}$ explicitly.
In the case $n=2$ the resulting equation for the two-body $t$-matrix
has the same formal structure as the Feynman-Galitzkii equation
\begin{equation}
T_2=K+ K R^{(2)}_0 N_2  T_2
\label{eqn:T2}
\end{equation}
where $K$ represents the two-body interaction kernel.
For $n=3$ the starting point is also given by the Bethe-Salpeter equation
\begin{equation}
T_3=K+ K R^{(3)}_0 N_3  T_3.
\label{eqn:T3}
\end{equation}
A derivation of relativistic three-body equation on the light-front is
formally identical to the
nonrelativistic~\cite{Beyer:1999zx,Beyer:2000ds} case, if we
neglecting antiparticle degrees of
freedom. Following~\cite{Beyer:1999zx,Beyer:2000ds,Mattiello:Diss} the
resulting equation for the vertex function $\ket{\gC_\ga}$ reads
\begin{equation}
\label{eqn:Fad-Vertex}
\ket{\gC_\ga}=\sum^{3}_{\gb=1}(1-\gd_{\ga\gb})N_2^\gb T_2^\gb R^{(0)}_ {2,\gb}\ket{\gC_\gb},
\end{equation}
where $N_2^\gb$, $T_2^\gb$ and $R^{(0)}_ {2,\gb}$ indicate the Pauli
blocking factor, two-body $t$-matrix and the two-body resolvent in the
channel $\gb$.

%The Dyson approach allows us to consistently derive relativistic few-body equations for particles embedded in a medium of
%both finite temperature and finite density.

%We systematically include the effects of self energy corrections
%$m=m(T,\mu)$ and Pauli blocking factors, given in terms of the Fermi
%distribution function.

\section{NJL model at finite temperature and density on the light-front}
In this Section we apply the light-front finite temperature field theory to the
two flavour NJL model
\begin{equation}
\label{eqn:NJL}
{\mathcal L}=\bar\psi(i\gc_\mu{\partial}^\mu-m_0)\psi+G\left((\bar\psi\psi)^2+
(\bar\psi i \gamma_5\mbox{\boldmath{$\tau$}}\psi)^2\right),
\end{equation}
where $\mbox{\boldmath{$\tau$}}$ are the Pauli matrices.
%We show that it is possible to compute several phase boundaries
%known from instant form calculations \cite{Klevansky:qe} also in the framework
%of light front quantisation. These boundaries include the chiral restoration,
%the pion Mott dissociation, and the transition to colour superconductivity, see
%e.g.~\cite{Rischke:2003mt} for a review of the phase diagram of quark matter.
In the meanfield approximation the chiral quark condensate contributes to the dynamical generated
mass $m=m_0-2g\langle\bar\psi\psi\rangle$.
%One distinguishes the chiral
%symmetric and chiral broken phase by the condensate value
%$\langle\bar\psi\psi\rangle$.
%The light-front in-medium propagator relevant
%here has been given in \eqref{eqn:lfmed}.
Using (\ref{eqn:lfmed_long}) to compute the chiral quark condensate
leads to the in-medium gap equation
\begin{equation}
\label{eqn:in-mediumgapeqn}
m(T,\mu)=m_0+24G\int\frac{dk^+ d^2\vec{k}_\perp}{k^+(2\pi)^3}
m(1-f^+(k^+,\vec{k}_\perp)-f^-(k^+,\vec{k}_\perp)).
\end{equation}
The medium independent first summand in the integral in (\ref{eqn:in-mediumgapeqn})
needs to be regularised, as usual the temperature modifications to the
mass $m(T,\mu)$ are finite.  We like to use a consistent
regularisation scheme for the gap equation and the related two-body
calculations.  Therefore the two-body bound state problem shall be
discussed before treating the issue of regularisation of
(\ref{eqn:in-mediumgapeqn}). The dependence of the constituent quark
mass on the medium has been calculated within the NJL
model on the light-front in Ref.~\cite{Beyer:2005rd}.

\subsection{Pions}
\label{Sec:Pion}
The pion $t$-matrix $T_\pi(k)$ is obtained by the equation (\ref{eqn:T2}) that represented in momentum space reads
\begin{equation}
\label{eqn:BS_eqn}
T_\pi(k)=K+\int\frac{d^4q}{(2\pi)^4}KG(q+k/2)G(q-k/2)T_\pi(k),
\end{equation}
where $K$ is some irreducible interaction kernel and
$G(q)$ the quark propagator taking into account the thermal mass (\ref{eqn:in-mediumgapeqn}).
To arrive at (\ref{eqn:BS_eqn}) we replaced the two-body Green function $G^{(2)}_{0}(q)$ by $G(q-k/2)G(q+k/2)$
consistent with (\ref{eqn:Gn}).

%Using $G^{(2)}_{0}(q)=G(q-k/2)G(q+k/2)$ in equation~(\ref{eqn:T2})
%For the two-body $t$-matrix $T_\pi(k)$  the Bethe-Salpeter equation given in eq.(\ref{eqn:T2}) in momentum space reads
%\begin{equation}
%\label{eqn:BS_eqn1}
%T_\pi(k)=K+\int\frac{d^4q}{(2\pi)^4}KG^{(2)}_{0}T_\pi(k),
%\end{equation}
%holds,
%or equivalently
%\begin{equation}
%\label{eqn:BS_eqn}
%T_\pi(k)=K+\int\frac{d^4q}{(2\pi)^4}KG(q+k/2)G(q-k/2)T_\pi(k),
%\end{equation}
%where $K$ is some irreducible interaction kernel and
%$G(q)$ the quark propagator taking into account the thermal mass (\ref{eqn:in-mediumgapeqn}).
The separable interaction kernel in the pseudo-scalar channel reads
\begin{equation}
\label{eqn:pseudo-scalarChannel}
K_{\alpha\beta,\gamma\delta}=-2iG(\gamma_5\tau_i)_{\alpha\beta}(\gamma_5\tau_i)_{\gamma\delta}.
\end{equation}
Introducing a reduced $t$-matrix $t_\pi(k)$ via
$T_\pi(k)_{\ga\gb\gc\gd}=(\gc_5\tau_j)_{\ga\gb}t_\pi(k)(\gc_5\tau_j)_{\gc\gd}$
the solution to (\ref{eqn:BS_eqn}) is
\begin{equation}
\label{eqn:solutionofBS}
t_\pi(k)=\frac{-2iG_\pi}{1+2G_\pi\Pi_\pi(k^2)},
\end{equation}
where $\Pi_\pi(k^2)$ denotes the contribution of the bubble diagram for the pion.
Evaluation of $\Pi_\pi(k^2)$ using (\ref{eqn:Gn}) (or equivalently (\ref{eqn:lfmed_long})) leads to
\begin{equation}
\label{eqn:pion_inmediumloop}
\Pi_\pi(k^2)=-6\int\frac{dxd^2\vec{q}_\perp}{x(1-x)(2\pi)^3}\frac{M_{2|0}^2(x,\vec{q}_\perp)
\left(1-f^+(M_{2|0})-f^-(M_{2|0})\right)}{M_{2|0}^2(x,\vec{q}_\perp)-k^2},
\end{equation}
which depends on the mass of the virtual quark-antiquark system
$M_{2|0}^2(x,\vec{q}_\perp)=(\vec{q}_\perp^2+m^2)/x(1-x)$. We introduced
the longitudinal momentum fraction $x=q^+/k^+$.
The momentum dependence of the Fermi blocking factors can also be written in terms of $M_{2|0}$ as
\begin{equation}
\label{eqn:fermiblocking}
f^\pm(M_{2|0})=\left[\exp\left\{\gb\left(\frac{M_{2|0}}{2}\mp\mu\right)\right\}+1\right]^{-1}.
\end{equation}
The thermal mass of the pion $m_\pi(T,\mu)$ is determined by the pole
of $t_\pi(k)$.

The integrals involved in (\ref{eqn:in-mediumgapeqn}) and
(\ref{eqn:pion_inmediumloop}) are divergent and need regularisation.
We utilise the invariant Lepage-Brodsky
(LB) cut-off scheme, which restricts the mass $M_{2|0}$ by
\begin{equation}
\label{eqn:LBscheme}
M_{2|0}^2(x,\vec{q}_\perp)=\frac{\vec{q}_\perp^2+m^2}{x(1-x)}\leq\gL_{\rm LB}^2.
\end{equation}

We are left with three model parameters, namely the quark current mass
$m_0$, the coupling constant $G$, and the cut-off $\gL_{\rm LB}$. These
are adjusted such that the pion mass $m_\pi=140$ MeV, the pion
decay constant $f_\pi=93$ MeV, and constituent quark mass $m=336$ MeV
are obtained correctly. One finds the values
\begin{eqnarray}
\label{eqn:parameters}
G&= 5.51\; {\rm eV},\\
\nonumber
m_0&= 5.67\; {\rm MeV},\\
\nonumber
\Lambda_{\rm LB}&= 1428\;{\rm MeV}.
\end{eqnarray}
These correspond to the case I discussed in~\cite{Klevansky:qe}.  Following
Ref.~\cite{Bentz} the gap equation with LB regularisation in
light-front form and the gap equation with 3-momentum (3M) cut-off
($\vec{k}^2\leq \gL_{\rm 3M}^2$) in instant form are equivalent if one
chooses a medium dependent LB cut-off
\begin{equation}
\label{eqn:LB-3M}
\gL_{\rm LB}^2(T,\mu)=4\left(\gL_{3M}^2+m^2(T,\mu)\right).
\end{equation}
The medium dependence of the cut-off may seem artificial at first, but
one should notice that the relevant energy scale of the effective
field theory changes with the medium parameters.

The in-medium pion mass for different values of the chemical potential
is shown in Figure~1.A. The Mott dissociation line is given by the
intersection of $m_\pi(T,\mu)$ and the continuum $2m(T,\mu)$. In the
NJL model discussed the chiral phase transition occurs when
$m(T_c,\mu_c)=m(0,0)/2$ holds~\cite{Klevansky:qe}. In Figure~3.B the
chiral phase transition $(T_c,\mu_c)$ is plotted as solid line and one
reads off the critical temperature for vanishing chemical potential as
$T_c=190$ MeV which is compatible with recent lattice calculations with almost physical quark masses~\cite{Cheng:2007jq}. The
dashed line represents the pion dissociation. It worth noting that
the pion dissociation and the chiral phase transition line are located
in a narrow band of the phase diagram, which is of course expected by
Goldstone boson character of the pion. However, both lines do not
coincide in this simple model of quark matter, which means that
deconfinement and chiral restoration happen at different temperatures
and densities in accordance with certain lattice
computations~\cite{Aoki:2006br}.
\begin{center}
\begin{figure}[t]
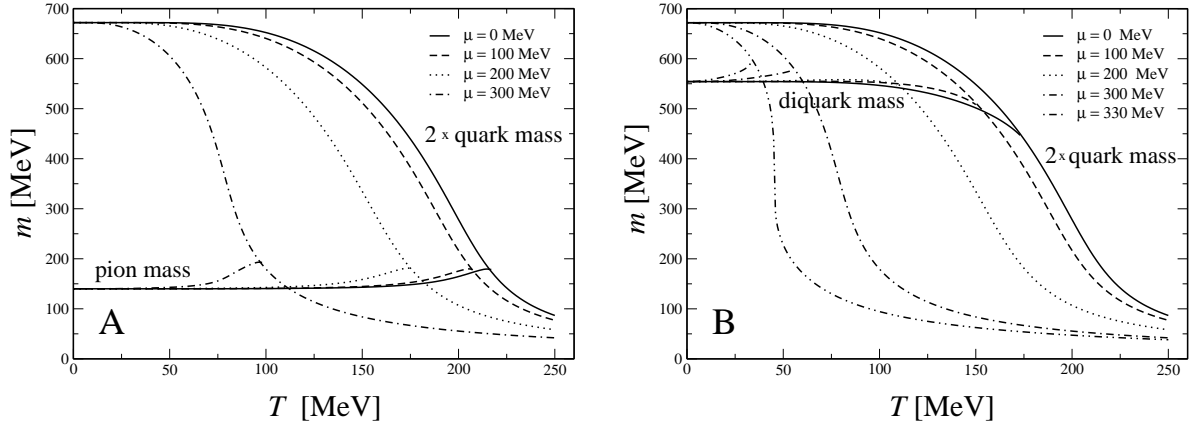

\begin{minipage}[t]{0.48\textwidth}
\includegraphics[width=\textwidth]{pionmass.eps}
\end{minipage}%
\begin{minipage}{0.04\textwidth}
   \hfill %
\end{minipage}%
\begin{minipage}[t]{0.48\textwidth}
\includegraphics[width=\textwidth]{diquark.eps}
\label{fig1}
\end{minipage}
\caption{
(A) The pion mass as a function of $T$ for different $\mu$.
(B) The diquark mass as a function of $T$ for different $\mu$.
The continuum is given by $2m$. The lines of
the two-body masses end at the Mott dissociation points.}
\end{figure}
\end{center}

\subsection{Diquarks and Colour Superconductivity}
The quark-quark interaction is constructed by Fierz transformation of
(\ref{eqn:NJL}), see Reference~\cite{Ishii:1995bu} for details.
Let us consider scalar, isospin singulet, colour-antitriplet diquarks, i.e.
we choose the following kernel
\begin{equation}
\label{eqn:diquark_kernel}
K_{\alpha\beta,\gamma\delta}=2iG_s(\gamma_5C\tau_2\gl^a)_{\alpha\beta}
(C^{-1}\gamma_5\tau_2\gl^a)_{\gamma\delta},
\end{equation}
with $\ga=2,5,7$ for the anti-symmetric Gell-Mann matrices in standard
representation and the charge conjugation matrix $C=i\gc^2\gc^0$.
Analogous to the pion case the reduced $t$-matrix $t_d(k)$ is
introduced and the solution
\begin{equation}
\label{eqn:BS_diquark_solution}
t_d(k)=\frac{2iG_d}{1+2G_d\Pi_d(k^2)},
\end{equation}
contains the diquark loop
\begin{equation}
\label{eqn:diquark_inmediumloop}
\Pi_d(k^2)=-3\int\limits_{\rm LB}\frac{dxd^2\vec{q}_\perp}{x(1-x)(2\pi)^3}\frac{M_{2|0}^2(x,\vec{q}_\perp)
\left(1-2f^+(M_{2|0})\right)}{M_{2|0}^2(x,\vec{q}_\perp)-k^2},
\end{equation}
where the normalisation factor~(\ref{eqn:NormDiq}) is present.

Results for the medium dependence of the diquark mass $m_d$ are
presented in Figure 1.B.  The coupling in the diquark interaction
channel is fixed to $G_d/G_\pi=1/\sqrt{2}$ which leads to the mass
$m_d\simeq 554$ MeV of an isolated diquark. Interestingly the
dependence of the in-medium diquark mass on temperature changes
drastically when the chemical potential is increased. While for small
$\mu$ the diquark mass decreases with temperature, the behavior is
opposite for $\mu\gtrsim 200$ MeV. For even larger chemical potentials
the chiral phase transition occurs and the diquarks cease to exist as
bound states. This turn-around is caused by the competition of two
effects, namely Pauli blocking and the in-medium quark mass. The contribution to the diquark loop
free from Pauli blockings is still medium dependent and leads to a decreasing
bound state mass as the quark mass reduces. Due to the different couplings in the pion and
diquark channel one observes this differing behavior of the bound
state mass even when the loop integrals (\ref{eqn:pion_inmediumloop})
and (\ref{eqn:diquark_inmediumloop}) are equal, e.g. for $\mu=0$.

The interaction in the above discussed quark-quark channel is
attractive. Therefore one expects the formation of a colour superconducting phase for
sufficiently high densities and low temperatures. For two flavours the
colour superconducting phase (2SC) is characterised by the condensate
$\Phi=\langle\psi^TC\gc_5\tau_2\gl_2\psi\rangle$,
cf. \cite{Buballa:2003qv} for a review. The SU$_c$(3) colour symmetry is broken
down to SU(2). Usually one derives a gap equation for $\Phi$ but here we
follow a different strategy  using the diquark $t$-matrix.
The Thouless criterion
\begin{equation}
\label{eqn:Thoulesscriterion}
m_d(T,\mu)=2\mu
\end{equation}
enables us to compute the boundary to the colour superconducting
phase~\cite{Thouless:AP10}. Equation (\ref{eqn:Thoulesscriterion}) is a condition for the condensation of bosonic diquarks but remains valid beyond the two
quark threshold and leads to the cancelation of the pole present in
(\ref{eqn:diquark_inmediumloop}) for $m_d\ge 2m$.  Inserting
(\ref{eqn:Thoulesscriterion}) into (\ref{eqn:BS_diquark_solution}) the
pole condition becomes
\begin{equation}
\label{eqn:CSCequation}
\frac{1}{2G_d}=3\int\limits_{\rm LB}\frac{dx}{x(1-x)}
\int\limits_{\rm LB}\frac{d^2\vec{q}_\perp}{(2\pi)^3}
\frac{M_{2|0}^2(x,\vec{q}_\perp)
\left(1-2f^+(M_{2|0})\right)}{M_{2|0}^2(x,\vec{q}_\perp)-4\mu^2}.
\end{equation}
Our result is plotted in quark matter phase diagram Figure~3.B.  The
2SC phase is located at temperatures below $T\le 92$ MeV and for
chemical potentials $300\;\rm{MeV}\lesssim\mu\le 702$ MeV. At $T=85$
MeV and $\mu=297$ MeV the boundary of the superconducting phase and
the chiral phase transition meet. These results are in general
agreement with specific scenarios of more sophisticated instant form
calculations including various condensates and colour superconducting
phases~\cite{Buballa:2003qv}.

\subsection{Nucleon}
On the light-front the treatment of the spin is technically involved
and for the time being we average over the spin projections.  This
procedure, used before in Ref.~\cite{deAraujo:1995mh} to describe the
proton electric form factor with a zero-range interaction, can be
justified in quark matter, because the spins can be regarded as washed out by
the medium.  Equation (\ref{eqn:Fad-Vertex}) evaluated in momentum space
for a zero-range interaction reads
\begin{equation}\label{eqn:3-body4k}
\gC(q) = \frac{2t(P_2)}{(2\pi)^4}\int d^4k R^{(0)}_1(k)N_2(k,P_3-q-k)R^{(0)}_2(P_3-q-k)\gC(k),
\end{equation}
where $R^{(0)}_n$ are $n$-body free boson propagators, $P_3^\mu$ is the three-body energy momentum vector in the center of mass system and the energy momentum vector of the two-body subsystem is given by $P_2^\mu=P_3^\mu-q^\mu$.
The two-body scattering amplitude $t(P_2)$ reads
\begin{equation}
t(P_2)=\left(i\lambda^{-1} - B(P_2)\right)^{-1},
\label{eqn:tau2med}
\end{equation}
where the loop integral is given by
\begin{equation}
B(P_2)=-\frac{i}{2(2\pi)^3} \int \frac{dx d^2k_\perp}{x(1-x)}
\frac{1-2f^+(x,\vec k^2_\perp)}{P_2^2-M_{2|0}^2}.
\label{eqn:Bmed}
\end{equation}
One notes the similarity of (\ref{eqn:diquark_inmediumloop}) and
(\ref{eqn:Bmed}) but the former integral accounts for the spins of the
quarks which then lead to the $M^2_{2|0}(x,\vec{q}_\perp)$ factor in the
numerator. After the integration of (\ref{eqn:3-body4k}) the
integral equation for the vertex function can be written as
\begin{eqnarray}
\Gamma(y,\vec q_\perp) = \frac{i}{(2\pi)^3} t(P_2) & \int_{0}^{1-y}\frac{dx}{x(1-y-x)} \\
\nonumber 
 &\;\; \int d^2k_\perp \frac{1-F(x,y;\vec k_\perp,\vec q_\perp)}{M^2_3 -M_{3|0}^2}\;\Gamma(x,\vec k_\perp),
\label{eqn:fad}
\end{eqnarray}
where the arguments of the blocking factors read
\begin{equation}
F(x,y;\vec k_\perp,\vec q_\perp)=f(x,\vec k^2_\perp)
+f(1-x-y,(\vec{k}_\perp + \vec{q}_\perp)^2).
\label{eqn:block}
\end{equation}
The integral involving $B(P_2)$ has a logarithmic divergence.
In order to investigate a solution of the
three-body bound state equation, we now extend the LB-regularisation
explained in Section~\ref{Sec:Pion} to the mass of the virtual
three-particle state $M_{3|0}$, which is the sum of the on-shell
minus-components of the three particles, i.e.
\begin{equation}
M^2_{2|0},M^2_{3|0}<\Lambda^2.
\end{equation}
It is not meaningful to use the same value of the cut-off used in the
two-body calculations, because the mass of the virtual
three-particle state should be $M_{3|0}\geq3m$ and in this case we
should have a too small integration range.  Therefore, we express the
parameter in units of the quark mass
following~\cite{Mattiello:TN03,stab:2003}, i.e. $\Lambda=\nu m$.  The
regularised two-body propagator is then modified as follows
\begin{equation}
B_\gL(P_2)=-\frac{i}{2(2\pi)^3} \int\limits_{M_{2|0}^2\le\gL^2} \frac{dx d^2k_\perp}{x(1-x)}
\frac{1-2f^+(x,\vec k^2_\perp)}{P_2^2-M_{2|0}^2}
\label{eqn:BmedREG}
\end{equation}
and the three-quark equation becomes~\cite{Mattiello:TN03}
\begin{eqnarray}
\label{eqn:med3}
\Gamma_\Lambda(y,\vec q_\perp) &= &\frac{i}{(2\pi)^3}\ t_\Lambda(M_2)
\int_0^{1-y} \frac{dx}{x(1-y-x)}\\
&&\int\limits_{M_{3|0}^2\le\gL^2} d^2k_\perp
\frac{%\theta(M^2_{30}-\Lambda^2)
1-f^+(x,\vec k_\perp)-f^+(y,(\vec k+\vec q)_\perp )}
{M^2_3 -M_{3|0}^2}\;\Gamma_\Lambda(x,\vec k_\perp)\nonumber.
\end{eqnarray}

\begin{center}
\begin{figure}[t]
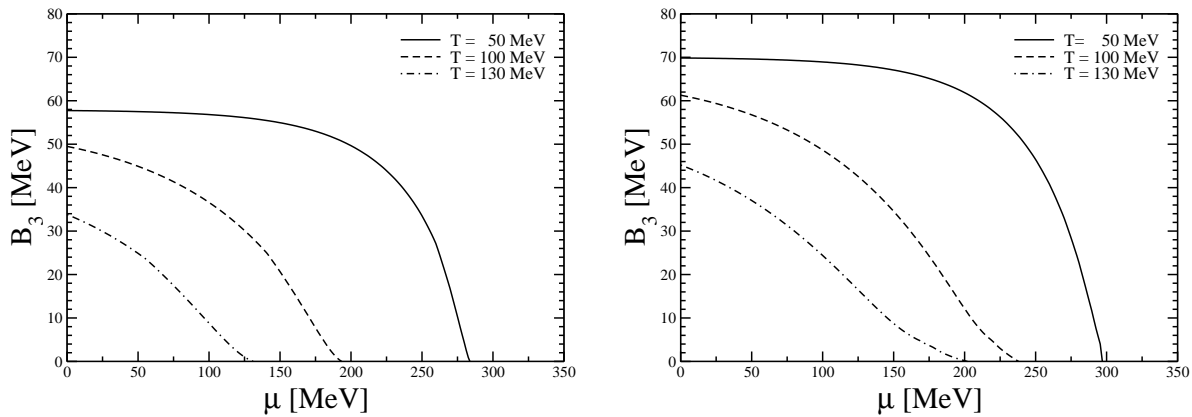

\begin{minipage}[t]{0.48\textwidth}
\includegraphics[width=\textwidth]{B3muLB04.eps}
\end{minipage}%
\begin{minipage}{0.04\textwidth}
   \hfill %
\end{minipage}%
\begin{minipage}[t]{0.48\textwidth}
\includegraphics[width=\textwidth]{B3muLBIF.eps}
\label{fig:BEmuLB}
\end{minipage}
\caption{
Binding energy of the three-quark bound state at various
temperatures as indicated for $\Lambda=4m$ (left panel) and for
$\Lambda=10^{15}m$ (right panel)}
\end{figure}
\end{center}

We have investigated the isolated two- and three-body bound states as
a function of the coupling $\lambda$ for different values of the
cut-off parameter and the stability of the three-body problem
elsewhere~\cite{Mattiello:TN03,stab:2003}.  Choosing $m=336$ MeV for
the quark mass we constrain our results to the nucleon mass. This determines
the function $\lambda(\Lambda)$ and the model is
parameterised by the cut-off $\Lambda$ only. In the following we use the
two limiting cases $\Lambda=4m$ and $\Lambda=10^{15}m$.

The solution of (\ref{eqn:med3}) allows us to calculate the
three-quark binding energy at finite temperatures and the chemical
potentials for the different cut-offs defined as
\be
B_3(T,\mu)=m(T,\mu)+M_{2B}(T,\mu)-M_{3B}(T,\mu).
\ee
In Figure 2. the binding energy $B_3$ as function of the chemical
potential for constant values of the temperature using $\Lambda=4m$
and $\Lambda=10^{15}m$ is shown.  $B_3$ becomes smaller by increasing
chemical potential for a constant value of the temperature.  We can
calculate at which temperature and chemical potential the binding
energy vanishes and therefore the three-quark bound states
disappear. As before the values of $T$ and $\mu$, for which the nucleon
dissociation occurs, define the Mott lines and are shown on the left
panel of Figure 3. in comparison with the chiral phase transition.  At low
temperatures the dependence on the cut-off is mild, but at zero
density the different cut-offs $\Lambda$ give significantly different
values for the transition temperature. This can be understood by
considering the important role of the colour screening at high density
which cancels the details of the dynamics given by the different
values of the cut-off. At low densities the effect of the colour
screening is negligible and the different dynamics are perceivable.
The Mott lines qualitatively follow the chiral phase transition given
by the solid line and the nucleon dissociation occurs in the chiral
broken phase. An overview of the phase diagram is given on the right
side of Figure 3. The interval between these two transitions may be
regarded as the confinement-deconfinement region.

\begin{center}
\begin{figure}[t]
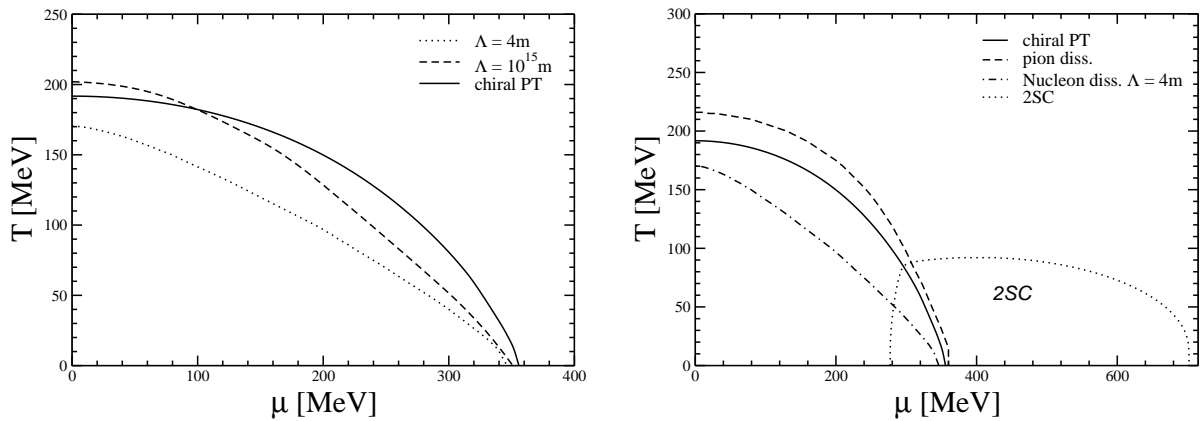

\begin{minipage}[t]{0.488\textwidth}
\includegraphics[width=\textwidth]{Mott3B.eps}
\label{fig:Mott3B}
\end{minipage}%
\begin{minipage}{0.04\textwidth}
   \hfill %
\end{minipage}%
\begin{minipage}[t]{0.48\textwidth}
\includegraphics[width=\textwidth]{phase.eps}
\label{fig:phase}
\end{minipage}
\caption{
(left panel) The Mott lines of the nucleon for $\gL=4m$ and
$\gL=10^{15}m$.  (right panel) The quark matter phase diagram in the light cone
NJL model. The solid line shows the chiral phase transition. The
dashed line is the Mott dissociation line of the pion and the
dashed-dotted line is the Mott dissociation of the nucleon for
$\gL=4m$. Finally the transition between the deconfined and the 2SC
phase is given by the dotted line.}
\end{figure}
\end{center}

\section{Conclusions}
Starting from a quasi-particle concept of quarks embedded in a hot and
dense medium, we presented a unified description of the most important
quark correlations present at the phase transition of QCD. Two
challenges have been faced, correlations in a many-particle system and
relativity. Correlations within a quasi-particle picture lead, in the
simplest case, to (properly modified) few-body equations for two- and
three-body states. Relativity leads to kinematical differences and the
possibility of pair creation. In addition the strength of the
interaction leads to a strong modification of the self energy of the
particles. These challenges have been tackled utilising the light-front
form of relativity.

The basically Hamiltonian concept of the light-front form allows us to
carry the Hamiltonian formulation of nonequilibrium quantum statistics
along with the Dyson expansion (or cluster expansion) of many-body
Greens functions through to the relativistic regime. For the sake of a
unified description we have utilised a simple, but rather useful and
hence widely investigated model of QCD, the Nambu--Jona-Lasino
model. The basic ingredients of QCD relevant for finite temperature
studies are included except for the confinement property. It is
reasonable to assume confinement to be screened strong enough due to
the presence of the other charged sources in the plasma. In fact, the
major quality of this interaction is the zero-range
property. Within this context we determine the in-medium properties of
pions, diquarks and nucleons and investigated the whole
temperature-density plane of the quark matter phase diagram including
the chiral phase transition, the onset of the colour superconducting
phase and the dissociation lines of pions and nucleons. The later ones
give a measure were the residual interaction between the quarks is weak
enough to allow the onset of the confinement-deconfinement phase
transition (assuming the colour dependent confining mechanism already
screened accordingly).

The physics field that can now be tackled is quite broad ranging from
such intriguing questions how hadrons form during the early plasma
phase of the universe, whether a colour superconducting phase exists,
up to what the nature of QCD phase transition is. In addition, going beyond
the quasi-particle approach seems to be straightforward in the Dyson
expansion, however, it is still intriguing to arrive at self consistent
results. Along this line a next step is the implementation of the full three-body
$t$-matrix in a computation of the critical temperature of
condensation/pairing of diquarks. The present formalism is well suited
to address this very interesting problem, hardly investigated in literature.

Besides the Dyson expansion in an equilibrated quasi-particle gas the
light-front formulation allows us also to derive relativistic
transport equations. First calculations that go beyond the NJL model
have already been performed in 1+1 dimensions.  Further effort is
needed to approach the full 3+1 domain utilising e.g. transverse lattice
methods or alike.

\end{document}